\documentclass[10pt]{article}
\usepackage[letterpaper,margin=1in]{geometry}
\usepackage{times}
\usepackage{amsmath,amssymb,booktabs,tabularx,array,graphicx,float}
\usepackage[round,authoryear]{natbib}
\usepackage{xcolor}
\definecolor{DCPBlue}{HTML}{176BE6}
\definecolor{DCPLinkBorder}{HTML}{D9E7FA}
\definecolor{DCPLinkBackground}{HTML}{F5F9FF}
\definecolor{DCPGitHub}{HTML}{24292F}
\usepackage{tikz}
\tikzset{
  DCPBadge/.style={rounded corners=5pt,inner xsep=8pt,inner ysep=4.5pt,font=\sffamily\small,line width=0.45pt},
  DCPPyPIBadge/.style={DCPBadge,draw=DCPLinkBorder,fill=DCPLinkBackground,text=DCPBlue},
  DCPGitHubBadge/.style={DCPBadge,draw=DCPGitHub,fill=DCPGitHub,text=white}}
\usepackage{hyperref}
\hypersetup{colorlinks=true,linkcolor=black,citecolor=black,urlcolor=DCPBlue,
  pdftitle={Scores Alone Do Not Prove Discovery: The Discovery Certification Protocol for Auditing AI Research Agents},
  pdfauthor={Jingjie Ning, Shanshan Zhong, Xiaochuan Li, Ji Zeng},
  pdfsubject={An executable evidence protocol for AI research agents},
  pdfkeywords={AI research agents, automated research, automated scientific discovery, discovery certification, outcome recovery, feedback evaluation}}
\usepackage{microtype}
\usepackage{amsmath,amsfonts,bm}

\def\eqref#1{equation~\ref{#1}}

\def\1{\bm{1}}

\DeclareMathAlphabet{\mathsfit}{\encodingdefault}{\sfdefault}{m}{sl}
\SetMathAlphabet{\mathsfit}{bold}{\encodingdefault}{\sfdefault}{bx}{n}

\newcommand{\astar}{A^\star}
\newcommand{\lineage}{L^\star}
\newcommand{\epsilonv}{\varepsilon}

\newcommand{\dcpverifierpackage}{\texttt{dcp-audit}}
\newcommand{\dcpharnesspackage}{\texttt{dcp-harness}}
\title{Scores Alone Do Not Prove Discovery:\\ The Discovery Certification Protocol for\\ Auditing AI Research Agents}
\author{Jingjie Ning\thanks{Corresponding author. \href{mailto:jening@cs.cmu.edu}{\texttt{jening@cs.cmu.edu}}.}\quad Shanshan Zhong\quad Xiaochuan Li\quad Ji Zeng}
\date{}

\newcommand{\dcplinkbadge}[3][DCPPyPIBadge]{\href{#2}{\tikz[baseline=(badge.base)]{\node[#1] (badge) {\strut #3};}}}
\newcommand{\dcppypiicon}{\raisebox{-2.5pt}{\includegraphics[height=12pt]{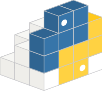}}}
\newcommand{\dcpgithubicon}{\raisebox{-2.5pt}{\includegraphics[height=12pt]{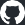}}}
\newcommand{\dcphomebadge}{\href{https://cxcscmu.github.io/Discovery-Certification-Protocol/}{\raisebox{-7.7pt}{\includegraphics[height=22pt]{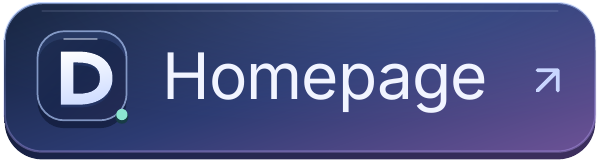}}}}
\makeatletter
\renewcommand{\maketitle}{%
  \begingroup
  \renewcommand{\thefootnote}{\fnsymbol{footnote}}
  \begin{center}
    {\fontsize{19}{23}\selectfont\bfseries\@title\par}
    \vspace{13pt}
    {\fontsize{12}{15}\selectfont\bfseries\@author\par}
    \vspace{6pt}
    {\normalsize School of Computer Science, Carnegie Mellon University\par}
    \vspace{4pt}
    {\small\texttt{\{jening, szhong2, xiaochu4, jizeng\}@cs.cmu.edu}\par}
    \vspace{11pt}
    \dcphomebadge\hspace{8pt}%
    \dcplinkbadge{https://pypi.org/project/dcp-audit/}{\dcppypiicon\hspace{5pt}PyPI / dcp-audit}\hspace{8pt}%
    \dcplinkbadge{https://pypi.org/project/dcp-harness/}{\dcppypiicon\hspace{5pt}PyPI / dcp-harness}\hspace{8pt}%
    \dcplinkbadge[DCPGitHubBadge]{https://github.com/cxcscmu/Discovery-Certification-Protocol}{\dcpgithubicon\hspace{5pt}GitHub}\par
  \end{center}
  \@thanks
  \endgroup
  \setcounter{footnote}{0}
  \vspace{4pt}
}
\makeatother
\begin{document}

\maketitle
\suppressfloats[t]

\begin{abstract}
AI research agents combine public information and experimental feedback to produce measurable results. The Discovery Certification Protocol (DCP) turns an outcome claim into an executable audit under a registered model, information boundary, and budget. Gate 1 validates useful improvement. Gate 2 tests recovery by matched agents given the starting information and observed Web content, with run history and new measurements withheld. Core requires adequate registered controls, zero recoveries, and a finite-sample recovery bound. Optional Gate 3 compares truthful and neutral feedback from a shared checkpoint; Evidence adds a supported effect and a null-policy equivalence check. Controlled SQLite and virtual catalyst audits pass both decision kernels. On real-data response surfaces, Yacht and Ionosphere pass the Core kernel after zero recoveries in 96 attempts, with an upper bound of \(0.0468\). Each target combines ten observed utilities and six predictions into a 16-entry data product. Yacht scores \(0.7677\) on reconstruction of all 32 switch effects, with utility-prediction MAE \(0.0315\) on its six unmeasured configurations. Fresh truthful continuations recover the target level in \(9/30\) and \(16/30\) trials, respectively, separating achieved utility from process repeatability. A deterministic verifier reproduces these local decisions from frozen records.
\end{abstract}

\begin{figure}[t]
\centering
\includegraphics[width=\linewidth]{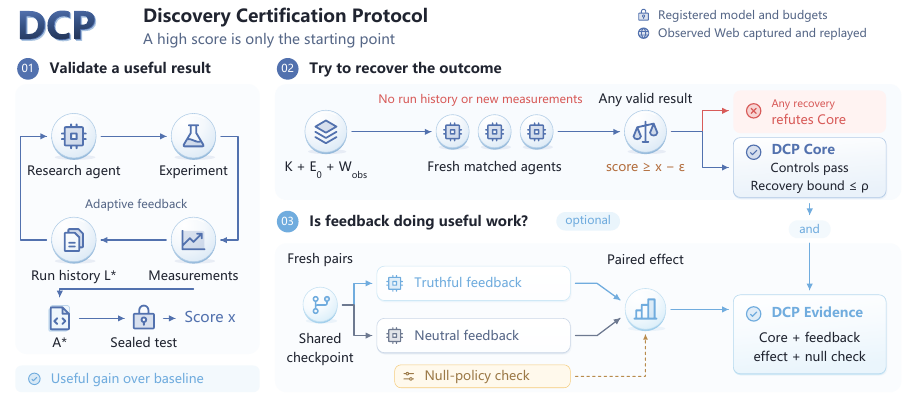}
\caption{DCP asks three questions about one numerical outcome. Gate 1 validates result $A^*$ on sealed data. Gate 2 gives fresh agents $K,E_0,W_{\mathrm{obs}}$ with run history and new measurements withheld. A valid score $\geq x-\varepsilon$ refutes Core; a Core pass requires adequate controls, zero recoveries, and a recovery bound below $\rho$. Gate 3 adds Evidence through a paired feedback effect and a null-policy equivalence check.}
\label{fig:teaser}
\end{figure}

\section{Introduction}

AI research agents choose actions, run experiments, inspect measurements, and revise their proposals. Their outputs include programs, models, data products, and experimental recipes. Recent systems have found useful algorithms and mathematical constructions and have begun to automate broader research workflows \citep{fawzi2022alphatensor,mankowitz2023alphadev,romera2024funsearch,novikov2025alphaevolve,lu2024aiscientist}. As these systems become research collaborators, their reports need evidence connecting measured results to the research that produced them.

Consider an agent that runs 87 experiments and reports a program with score \(x\). A sealed test establishes the program's utility. A matched agent may also reach \(x\) using initial data, public material, and its own reasoning. Separately, truthful experimental observations may improve the chance of reaching \(x\). These are three measurable questions about one outcome. Implementation and baseline choices shape measured gains \citep{melis2018evaluation,musgrave2020metric}. In automated research, independent implementations of a fixed idea can change its ranking \citep{ning2026implementation}. DCP extends this measurement discipline to the information and feedback available to a research agent.

The central design choice is to admit every valid route to the same numerical outcome. A challenger may combine familiar components, transfer a technique across domains, or produce a different implementation. It receives the registered background, initial observations, tools, budget, and captured Web content. Run history and new scientific measurements are withheld. An executable validity check and a score threshold determine recovery, making the rule portable across research domains.

Recovery witnesses and recovery probabilities play complementary roles. A witness establishes an available route within the registered scope and closes the audit as recovered. A positive Core decision requires an adequate audit with zero recoveries and a bound on a fresh episode's recovery probability. Evidence adds a randomized comparison of truthful and neutral feedback from a shared checkpoint. This separation gives a successful target run, a recoverable outcome, and a beneficial feedback policy distinct evidential meanings.

The paper makes three contributions.
\begin{itemize}
\item We define a general evidence interface for measurable AI research outputs. Its numerical recovery rule accepts alternative methods under a registered information and resource boundary.
\item We combine qualified recovery witnesses, finite-sample recovery bounds, and randomized feedback effects in two explicit decisions, Core and Evidence. Their definitions separate certificate eligibility from statistical and causal conclusions.
\item We demonstrate local Core and Evidence decisions in controlled workflows and two real-data Core passes with Evidence Inconclusive for distinct reasons. Recovery witnesses and incomplete records exercise the remaining decision paths. Portable bundles support deterministic replay.
\end{itemize}

\section{Related Work}

\paragraph{AI research agents and their evaluation.} AlphaTensor, AlphaDev, FunSearch, and AlphaEvolve combine search with objective evaluation \citep{fawzi2022alphatensor,mankowitz2023alphadev,romera2024funsearch,novikov2025alphaevolve}. The AI Scientist family and AI co-scientist automate broader research and hypothesis development \citep{lu2024aiscientist,yamada2025aiscientistv2,gottweis2025coscientist}. A specialist-agent ablation removes prior-trial history while retaining current-best code and score \citep{ning2026specialists}. Molecular and materials agents freeze selected interventions for held-out evaluation \citep{ning2026molecular,ning2026materials}. MLE-bench, MLGym, MLR-Bench, and PaperBench evaluate engineering, open-ended research, and replication \citep{chan2024mlebench,nathani2025mlgym,chen2025mlrbench,starace2025paperbench}. DiscoveryBench and ScienceAgentBench assess data-driven scientific workflows and their outputs \citep{majumder2025discoverybench,chen2025scienceagentbench}. DiscoveryWorld measures task success, scientific actions, and explanatory knowledge \citep{jansen2024discoveryworld}. FIRE-Bench evaluates rediscovery of published scientific insights \citep{wang2026firebench}, while \citet{bhushan2026creativity} distinguish psychological novelty, historical novelty, and usefulness. DCP adds outcome recovery and feedback interventions to these evaluations.

\paragraph{Iterative feedback and experimental design.} ReAct, Tree of Thoughts, Reflexion, and Self-Refine develop tool use, search, reflection, and feedback \citep{yao2023react,yao2023tree,shinn2023reflexion,madaan2023selfrefine}. Controlled revision studies separate additional solving, prompt structure, and draft content \citep{ning2026resolving}. SkillLearnBench compares execution-derived and teacher-guided skill refinement \citep{zhong2026skilllearnbench}. DCP tests feedback through randomized pairs \citep{rubin1974causal}, an equivalence region for independent null-policy checks \citep{schuirmann1987equivalence}, and preregistered confirmatory choices \citep{nosek2018preregistration}.

\paragraph{Evaluation integrity and process attestation.} Broad evaluation, contamination studies, and memorization research examine the influence of prior exposure \citep{hendrycks2021mmlu,srivastava2022bigbench,liang2022helm,golchin2023contamination,carlini2021extracting}. DeepResearchGym provides stable retrieval from fixed Web corpora \citep{coelho2026deepresearchgym}. SWE-bench emphasizes executable outputs \citep{jimenez2024swebench}. Proof-of-Learning records training states to attest a training procedure \citep{jia2021proofoflearning}. ScientistOne traces claims, scores, references, and code to supporting evidence \citep{meng2026scientistone}. DCP adds recovery and feedback interventions to research audits.

\section{Method}
\label{sec:method}

\subsection{The outcome and its information boundary}

An operational discovery claim specifies a useful outcome, a registered model and budget, a starting information packet, and a numerical recovery rule. Core bounds matched recovery with run history and new measurements withheld. Evidence estimates a feedback-policy effect from a registered checkpoint.

\(K\) contains fixed background, the model, tools, and known methods. \(E_0\) contains starting observations fixed independently of research actions. \(L^*\) is the target run's history used by later actions. \(A^*\) is the final output selected before sealed scoring, and \(P\) is the executable validity and recovery rule.

The \(E_0\) and \(\lineage\) boundary follows information provenance. An example present at the start belongs to \(E_0\). A measurement returned because the agent chose an action belongs to \(\lineage\). Fixed human guidance enters \(E_0\); guidance chosen after intermediate results enters a human-agent lineage. Core targets adaptive research with at least one new observation consumed by a later research action. A pool of independent candidates followed by final score-based selection is a search-and-selection baseline.

The validity predicate and numerical threshold define recovery across all admissible artifacts. Historical priority is a separate scholarly assessment. The same outcome audit applies to programs, models, data products, and recipes.

\subsection{Precommitment and sealed evaluation}

Registration precedes the target run and fixes the task, model, information, interface, tools, budgets, baseline, validity rules, selection and stopping procedures, and statistical analysis. It specifies the complete production procedure, including authorized measurement queries and their disclosure policy. Returned task measurements enter \(\lineage\); the evaluator protects the remaining private reference data and final outcome scores. Every started confirmatory audit consumes a preallocated error budget, including recovered and incomplete audits. The registered selection rule freezes \(\astar\) before final scores are revealed. Committed baseline, target, and control artifacts share the same validity and scoring rules.

\subsection{Gate 1 establishes useful improvement}

Let \(\operatorname{Valid}(a)\) denote compliance with the registered artifact constraints. Let \(b\) denote the baseline artifact and \(x\) the mean sealed score of \(\astar\), on a registered \([0,1]\) scale. Write \(\operatorname{LCB}\) for the registered lower confidence bound. For the smallest useful gain \(\delta_{\min}>0\), Gate 1 requires
\begin{equation}
    \operatorname{LCB}\!\left[
    \mu(\astar)-\mu(b)\right] \geq \delta_{\min},
    \label{eq:main}
\end{equation}
where \(\mu\) is mean utility over the registered evaluation population. Sampled populations use a finite-sample confidence interval; exhaustive finite evaluations use an exact mean. Gate 1 also requires valid baseline and target artifacts and a baseline below the recovery region. Confirmed target invalidity or insufficient utility fails the audit. Unresolved checks produce an inconclusive decision.

\subsection{Gate 2 tests recovery under a registered challenger}

Gate 2 gives a fresh matched agent the same \(K\), complete \(E_0\), model, interface, known components, and registered production opportunities. It withholds \(\lineage\) and gives the challenger all Web bytes observed by the target run, denoted \(W_{\mathrm{obs}}\). The agent may reason, compile, and perform engineering checks. Responses to its actions follow a frozen policy that supplies non-directional messages and withholds new task scores and scientific measurements.

Every valid method is eligible. For a registered tolerance \(\epsilonv\), the recovery rule is
\begin{equation}
    P(a) =
    \operatorname{Valid}(a)
    \ \wedge\
    \operatorname{score}(a) \geq x-\epsilonv .
    \label{eq:recovery}
\end{equation}
The tolerance expresses a small substantively equivalent score difference. Registration requires \(0\leq\epsilonv<\delta_{\min}\). Gate 1 checks the baseline's position after sealed scoring. Repeated paired evaluation and simultaneous confidence intervals handle measurement uncertainty. Generators receive the metric and selection rule; the verifier computes \(x\) after all artifacts are committed.

A qualified recovery witness satisfies both Equation~\ref{eq:recovery} and the registered execution, information, and provenance conditions. It provides an executable alternative route under the claimed resources and triggers the Core veto, recorded as \emph{recovered} with Core \emph{refuted}. The veto tests the existence of an audited alternative route. The probability bound below quantifies how often a fresh registered episode finds one. Both quantities remain in the report when a rare route is found.

\begin{figure}[!ht]
\centering
\includegraphics[width=\linewidth]{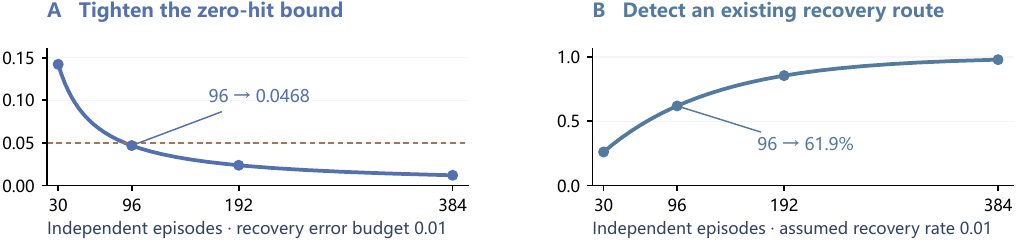}
\caption{Two effects of audit size. The zero-hit upper bound uses recovery error budget \(0.01\); the orange line marks \(\rho=0.05\). Witness detection assumes an illustrative recovery rate \(p_B=0.01\). At 96 episodes, the corresponding values are \(0.0468\) and \(61.9\%\).}
\label{fig:audit-sensitivity}
\end{figure}

Let \(Q_B\) be the registered distribution of complete challenger episodes at budget \(B\). An episode includes its model calls, candidate opportunities, and selection rule. A best-of-\(k\) procedure counts as one \(k\)-candidate episode. Let \(H=1\) when any admissible candidate recovers the target, and define \(p_B=\Pr_{Q_B}(H=1)\). Registration fixes audit size \(n\), recovery-bound threshold \(\rho\), and one-sided error allocation \(\alpha_{\mathrm{recovery}}\). With zero recoveries in \(n\) independent episodes, the exact Clopper--Pearson upper bound is \citep{clopper1934use}
\begin{equation}
    p_{\mathrm{upper}} =
    1-\alpha_{\mathrm{recovery}}^{1/n}.
    \label{eq:zero}
\end{equation}
Core requires zero qualified witnesses in the registered \(n\) episodes, \(p_{\mathrm{upper}}\leq\rho\), and complete audit checks. This supports \(p_B\leq\rho\) at confidence \(1-\alpha_{\mathrm{recovery}}\) under the stated sampling model. Checks cover opportunity budgets, information access, independent draws, valid-output rates, and the registered control task. Controls establish the operation they exercise; challenger scores measure search performance. Candidate score tests retain a separate error allocation. Core concerns the achieved level \(x\); fresh truthful continuations measure process repeatability. Figure~\ref{fig:audit-sensitivity} shows the distinct effects of audit size on precision and witness detection.

\subsection{Gate 3 estimates a feedback-policy effect}

\begin{figure}[!ht]
\centering
\includegraphics[width=\linewidth]{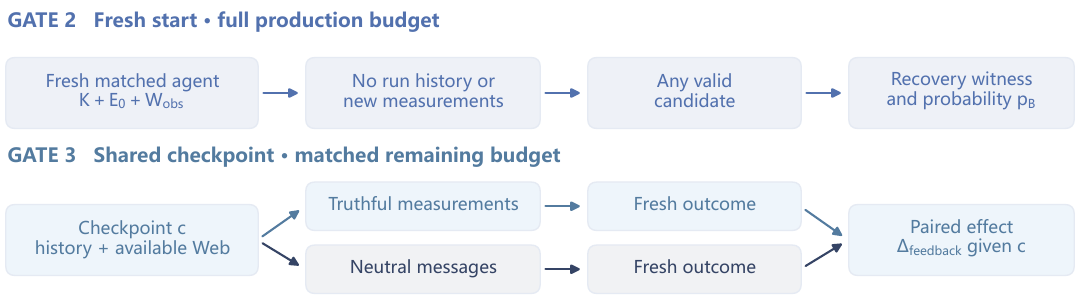}
\caption{Information flows in the two tests. Gate 2 starts with \(K,E_0,W_{\mathrm{obs}}\), withholding run history and new measurements. Gate 3 branches from one checkpoint containing its available history and Web, holds remaining resources fixed, and compares truthful and neutral feedback.}
\label{fig:gate-comparison}
\end{figure}

Gate 2 measures recovery from the starting information. Gate 3 measures how subsequent feedback changes fresh outcomes from a shared checkpoint \(c\). A rule registered before the run selects \(c\), including its files and available history. Fresh paired branches receive either truthful feedback from their own actions or messages from a specified neutral policy. Both arms share the model, tools, remaining budget, starting state, and output checks. Figure~\ref{fig:gate-comparison} summarizes both tests.

Registration specifies the timing, schema, and content of both feedback policies. The neutral policy withholds directional task measurements, and later actions may diverge in response to the assigned messages. Arm assignment, pair count, selection, and stopping are frozen before the study. Fresh replicates estimate the average policy effect conditional on the checkpoint.

For registered utility \(u\), the estimand is
\begin{equation}
    \Delta_{\mathrm{feedback}} =
    \mathbb{E}\!\left[
    u(A_{\mathrm{truthful}})-u(A_{\mathrm{neutral}})
    \mid c\right].
    \label{eq:feedback}
\end{equation}
A binary utility compares recovery probabilities. This is the total effect of the two registered feedback policies, conditional on \(c\), including their downstream influence on the agent's actions.

Independent null tasks compare the neutral policy with a second non-directional policy on problems whose answers are determined by \(E_0\). This null-policy equivalence check requires its interval to lie inside \(\pm\delta_{\mathrm{sham}}\). Its scope is the known-answer family. The target-task effect in Equation~\ref{eq:feedback} includes the behavior induced by the chosen neutral policy. The registered buffer \(\delta_{\mathrm{sham}}\) is a decision margin, with no assumed bound on target-task neutral-policy bias. Evidence requires \(\operatorname{LCB}(\Delta_{\mathrm{feedback}})\geq\delta_{\mathrm{evidence}}+\delta_{\mathrm{sham}}\) and the equivalence check. A frozen rule retains every start and governs pre-action pair replacements.

Appendix Figure~\ref{fig:sqlite-trace} traces the SQLite-Web audit through all three gates.

\subsection{Web access}

A recording gateway stores each query, response time, and exact model-visible bytes. Gate 2 discloses the complete observed packet at the start, so its recovery bound conditions on \(W_{\mathrm{obs}}\). Gate 3 pairs inherit the pages available at their checkpoint and share a frozen environment for later Web requests. A claim about the contribution of Web access uses an additional Web-withholding intervention.

\subsection{Threat model and trust boundary}

DCP treats the claimant and evidence producer as potentially strategic. In a formal deployment, an independent audit authority approves the question and resource match, holds the sealed test, and anchors records before execution. Independently anchored ledgers and provider-bound records expose missing starts, substituted models, early private-score access, and altered evidence. The checker evaluates numerical and record consistency under the accepted registration. Independent provenance verification and registry countersigning complete formal issuance.

Evidence production and decision checking have separate interfaces. Task adapters specify validity, scoring, and feedback; a shared deterministic verifier evaluates the resulting records. The reported audits were produced by task-specific runners. A lightweight reusable harness supports collection and control execution, and the offline verifier recomputes decisions from the resulting bundles. This separation lets independent readers check the same evidence across agent implementations.

\subsection{Decisions and their meaning}

Core and Evidence report \emph{Pass}, \emph{Refuted}, \emph{Inconclusive}, or \emph{Not evaluated} (Table~\ref{tab:decision-labels}). Inconclusive carries \emph{audit incomplete} or \emph{statistical uncertainty}. Prospective decisions use their pre-execution classification version. Table~\ref{tab:decision-provenance} identifies source policies and later amendments, with their timing and analysis scope. Appendix~\ref{app:failure-policy} defines these versioned failure rules.

\section{Experimental Audit Cases}
\label{sec:study}

\begin{figure}[!t]
\centering
\includegraphics[width=\linewidth]{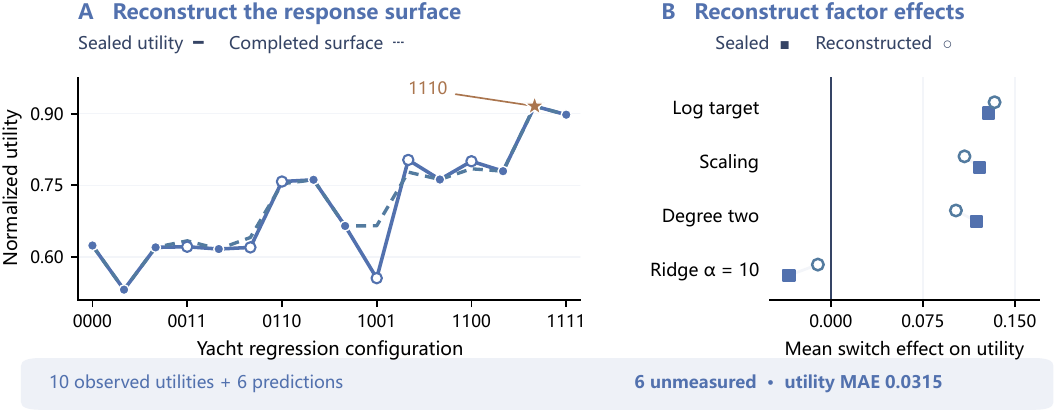}
\caption{The Yacht response surface combines ten observed utilities with six predictions. (A) Filled points mark the two public anchors and eight queried utilities inserted by the evaluator; open points mark undisclosed values. The star marks the accompanying pipeline selection. (B) Each mean effect averages eight matched pairs. The completed surface scores \(0.7677\) on all 32 effects; unmeasured-six utility-prediction MAE is \(0.0315\).}
\label{fig:real-findings}
\end{figure}

\subsection{Study design}

The primary suite combines Yacht and Ionosphere response-surface audits with controlled cases. Authors collected and replayed these records locally; every reported pass denotes a kernel decision with \path{formal_certificate_issued=false}. Independent authority review and countersigning define formal issuance. Artifacts were committed before final scoring under preallocated audit slots. Table~\ref{tab:decision-provenance} records version bases and original decisions; Appendix~\ref{app:dry-bean} presents an exploratory Dry Bean analysis.

Gate 1 enumerated each finite sealed workload. Gate 2 sampled fresh episodes from each registered \(Q_B\). SQLite-Web randomized equivalent presentations, catalyst randomized public starting recipes, and real-data tasks repeated fixed initial materials with provider-default sampling. Table~\ref{tab:episode-budgets} specifies these distributions and per-episode budgets \(B\); \(n=96\) is the audit sample size. Gate 3 sampled fresh paired continuations of a frozen checkpoint. Development cases guided neutral-channel design; policies, thresholds, equivalence bands, and pair distributions were frozen before confirmatory branches opened. Provider-reported identities were checked through the registered CLI. SQLite-Web used \texttt{deepseek-v4-flash}, catalyst used \texttt{deepseek-v4-pro}, and real-data sessions used \texttt{deepseek-flash} with matched interfaces and budgets.

\subsection{Real-data research workflows}

\begin{figure}[!t]
\centering
\includegraphics[width=\linewidth]{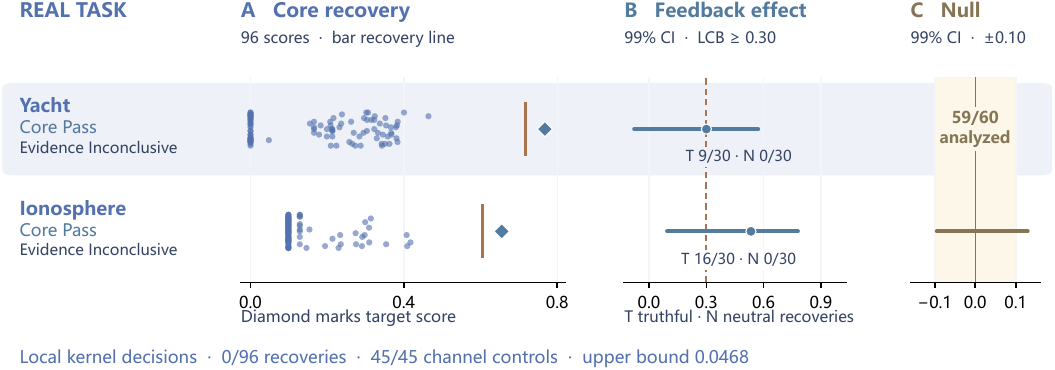}
\caption{Local Core passes and separate feedback decisions. (A) Challenger scores lie below the recovery lines; diamonds mark target scores. (B) T/N reports truthful/neutral recovery counts beside 99\% effect intervals; the required lower bound is \(0.30\). (C) The null band is \(\pm0.10\). Yacht analyzes 59 of 60 pairs; Ionosphere analyzes all 60. Both report Evidence Inconclusive, with reasons audit incomplete and statistical uncertainty, respectively.}
\label{fig:real-audit}
\end{figure}

Yacht uses hull and velocity data for regression \citep{gerritsma1981yacht}; Ionosphere uses radar returns for classification \citep{sigillito1989ionosphere}. These measurement-dependent profiles share device calibration's controlled information-access design, with utility surfaces obtained from pipelines on real data. Each scored artifact \(\astar\) combines two public anchors, eight measured utilities, and six predictions over four binary interventions. The evaluator inserts authorized observations into the submitted table. Baseline, challenger, and neutral observations contain only the two anchors; the baseline fills its other 14 entries with their mean. Gate 1 scores all 32 single-switch effects as \(\max(0,1-\mathrm{MAE}_{\mathrm{effect}}/\operatorname{mean}|\mathrm{effect}|)\). This evaluates the total acquisition-and-completion gain. Core recovery concerns this data product; selected-pipeline utility is an accompanying readout.

Yacht and Ionosphere vary transformations, scaling, complexity, and regularization. Each audit registered 96 challengers, 45 file-channel controls, 30 feedback pairs, and 60 null pairs. Controls copy a supplied marker and submit a finish action, testing the read/write channel. Gate 3 starts at checkpoint 0 before each branch selects eight masks. Truthful replies return their utilities; neutral replies contain \texttt{status=no-measurement} and an empty map. The end-to-end effect includes direct insertion and subsequent predictions. The neutral arm shares Gate 2's task state, prompt, budget, and response policy; Gate 2 additionally carries an empty-Web manifest. Separate sampling records preserve this implementation distinction.

\subsection{Controlled calibration profiles}

\begin{figure}[!t]
\centering
\includegraphics[width=\linewidth]{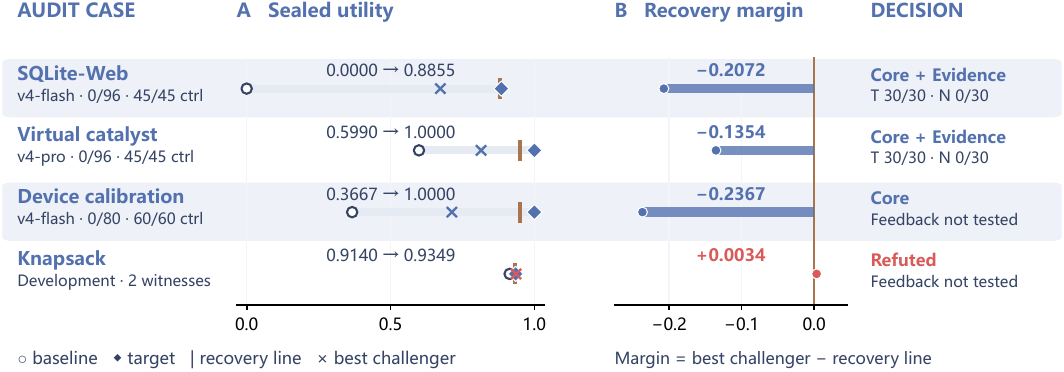}
\caption{Controlled audit outcomes. (A) Sealed baseline, target, recovery line, and best challenger. (B) Best-challenger score minus the recovery line; a positive margin supplies a witness. Recovery and control counts accompany each case. Knapsack supplies two developmental witnesses.}
\label{fig:controlled-decisions}
\end{figure}

The SQLite-Web audit in Figure~\ref{fig:controlled-decisions} presents 16 query families, four high-traffic families, and a four-index budget. Its score measures the reduction in traffic-weighted SQLite virtual-machine work while preserving every query result. The agent captured two documentation pages and used measured per-family work to select indexes. Gate 2 received the pages; Gate 3 compared measured traffic feedback with schema-matched uniform weights from a checkpoint after Web capture.

Virtual catalyst control offers \(8^5=32{,}768\) recipes. An agent selects an anchor, receives a 36-well one-factor plate, and commits a recipe evaluated on 256 sealed conditions. Truthful and neutral continuations share the anchor checkpoint. Device calibration supplies a Core control with known hidden gains. Multidimensional knapsack supplies two distinct legal recovery witnesses from one developmental challenger episode. An undersampled affine audit exercises incomplete control adequacy. Appendix~\ref{app:real-workflows} and the controlled audit records give the exact task and execution details.

\section{Results}

The main results connect response-surface reconstruction, matched recovery attempts, and feedback effects. Figures~\ref{fig:real-findings} and \ref{fig:real-audit} show the real-data evidence; Figure~\ref{fig:controlled-decisions} shows distinct controlled-case decisions.

\subsection{The real workflow reconstructs an informative response surface}

Yacht's completed surface scores \(0.7677\), against baseline \(0\), with utility-prediction MAE \(0.0315\) on its six unmeasured configurations. Its reconstructed mean utility effect of log-transforming the target is \(+0.1334\), against the sealed \(+0.1284\); scaling gives \(+0.1089\) against \(+0.1210\). Figure~\ref{fig:real-findings} also shows polynomial-feature and regularization effects. The surface selects measured configuration \texttt{1110}, combining target transformation, scaling, degree-two features, and Ridge \(\alpha=1\). Its utility \(0.9159\) accompanies the audited data product as a selection readout. The Core recovery line \(0.7177\) applies to effect reconstruction across the complete surface.

\subsection{Real-data audits separate recovery and feedback evidence}

Yacht and Ionosphere pass the Core kernel at outcome levels \(0.7677\) and \(0.6551\), with zero recoveries in 96 episodes and \(45/45\) channel controls. Their best challengers score \(0.4641\) and \(0.4171\), below recovery lines \(0.7177\) and \(0.6051\); the recovery upper bound is \(0.0468\). Fresh truthful continuations recover these levels in \(9/30\) and \(16/30\) trials; both neutral arms recover \(0/30\). Core evaluates the achieved level, while these rates quantify process repeatability.

Both Evidence decisions are Inconclusive. Yacht's feedback interval is \([-0.078,0.571]\), and a changed protected observation leaves its null-policy check audit incomplete. Ionosphere's feedback interval is \([0.094,0.779]\), and its null interval \([-0.095,0.129]\) crosses the equivalence band, yielding statistical uncertainty. Figure~\ref{fig:real-audit} presents these separate decision fields. The version table records the Ionosphere timeout classification and its unchanged final verdict.

\subsection{Controlled audits exercise the decision space}

\begin{figure}[!ht]
\centering
\includegraphics[width=\linewidth]{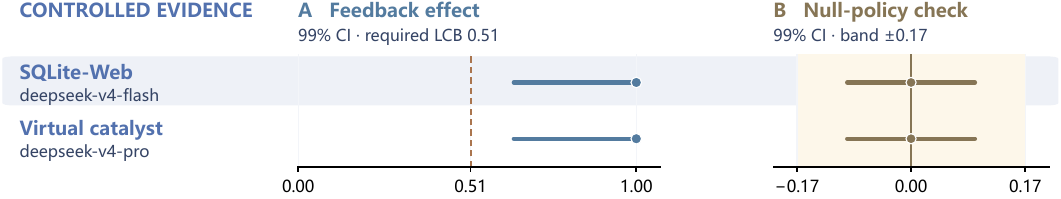}
\caption{Controlled Gate 3 evidence. Each audit has 30 target pairs and 60 null pairs. Registered 99\% feedback intervals clear the \(0.51\) lower-bound requirement; null intervals lie within \(\pm0.17\).}
\label{fig:controlled-feedback}
\end{figure}

SQLite-Web reduced traffic-weighted work by \(88.55\%\), and virtual catalyst control reached its sealed optimum. Both pass the Core and Evidence kernels (Figure~\ref{fig:controlled-feedback}). SQLite-Web's Evidence pass uses the corrected implementation of its registered read-before-first-write rule; Table~\ref{tab:decision-provenance} records the cross-bundle replay and version basis. Device calibration adds a Core pass, and knapsack supplies a recovery witness (Figure~\ref{fig:controlled-decisions}). The affine case and a separate SQLite addendum remain Inconclusive. Appendix Table~\ref{tab:controlled-values} gives exact values.

\section{Discussion}

\subsection{The recovery rule admits alternative methods}

A component ablation evaluates an intervention within a chosen implementation. DCP evaluates recovery across every admissible implementation offered by the registered challenger. A recombination of known modules, a transfer from another domain, or a different program can all supply the same numerical witness. Two distinct legal knapsack solutions cross the recovery line under this rule. Task adapters specify validity, sealed utility, and a starting information packet. Program search \citep{romera2024funsearch,novikov2025alphaevolve}, AI scientist workflows \citep{lu2024aiscientist,yamada2025aiscientistv2}, and ML engineering benchmarks \citep{chan2024mlebench,nathani2025mlgym} can instantiate the same audit interface when their candidate histories and budgets are registered.

\subsection{Reporting a research contribution}

A reusable report gives the outcome artifact, threshold, model and information scope, episode budget, recovery witness or bound, truthful repeatability, and checkpoint-conditional feedback interval. Authors submit the record, and auditors replay decisions and examine alternative routes. Yacht and Ionosphere illustrate local Core passes with distinct reasons for Evidence Inconclusive. Multi-task deployments register each task instance and report the decision distribution. A separately registered no-measurement arm can assess target-task sensitivity to the neutral policy.

\subsection{Deployment and retrospective recovery}

Prospective use starts with an approved registration and sealed evaluator, followed by evidence collection, control execution, offline replay, and independent countersigning. Auditors can stop at the first qualified recovery and reserve paired studies for feedback claims. Confirmatory collection used 419 catalyst sessions costing \(54.27\) USD and 504 SQLite-Web sessions costing \(61.13\) USD. These totals cover target, challengers, controls, feedback pairs, and null pairs. Generation incurs model cost; a reader replays the saved evidence offline.

Retrospective use reconstructs an outcome target and seeks a qualified recovery witness. An audit with pre-publication-equivalent model and information access can test the original recovery scope. An audit using post-publication knowledge measures present-day recoverability. Records identify this temporal scope and their analysis-policy version. Formal issuance requires prospective registration and complete evidence collection.

\section{Conclusion}

The Discovery Certification Protocol gives measurable AI research outputs an executable, layered evidence interface under a registered model, information boundary, and budget. Gate 1 establishes sealed utility. Gate 2 accepts any valid route to the numerical target with run history and new measurements withheld; Core requires adequate registered controls, zero witnesses, and a finite-sample bound at a fixed audit size. Gate 3 compares truthful and neutral feedback from a shared checkpoint; Evidence requires a positive effect above a registered margin and a null-policy equivalence check. A deterministic verifier replays local decisions; independent authority review and countersigning complete formal issuance.

The evidence covers controlled software optimization, virtual experimental control, and response-surface reconstruction on real datasets. SQLite-Web reduced traffic-weighted work by \(88.55\%\), and virtual catalyst control reached its sealed optimum; both pass the Core and Evidence kernels. Yacht and Ionosphere pass Core with zero recovery in 96 attempts and \(45/45\) channel controls. Their truthful repeatability rates of \(9/30\) and \(16/30\) accompany Evidence Inconclusive decisions. Yacht's completed surface combines ten observed utilities with six predictions, scores \(0.7677\) on effect reconstruction, and achieves unmeasured-six utility-prediction MAE \(0.0315\). Knapsack supplies an executable recovery witness. These records make achieved outputs, alternative routes, process repeatability, and feedback effects inspectable through one evidence interface.

\clearpage
\section*{Reproducibility Statement}

Frozen machine-readable bundles and evidence records support the numerical claims. The appendices report task construction, registrations, model and interface contracts, evaluation units, thresholds, statistics, budgets, costs, and identifiers. The anonymous supplement contains the controlled examples, three real-data decision bundles, versioned provenance receipts, and figure sources. It also contains the installable \dcpverifierpackage{}, task-specific evidence producers, and reusable \dcpharnesspackage{}. Decision replay is deterministic and offline. Fresh evidence production uses the model and environment specified in each registration. The harness exposes capture, challenge, paired feedback testing, finalization, and bundle generation as separate operations.

\section*{Ethics Statement}

AI research agents can accelerate useful discovery while increasing the need for clear and contestable evidence. DCP supports accountable reporting by binding each decision to a registered model, information boundary, budget, probability bound, and replayable record. Certificates should be presented with this scope, and captured Web data should follow applicable privacy, licensing, and security requirements. Deterministic replay gives authors, reviewers, and independent auditors a common record on which to examine a research claim.

\section*{AI Use Statement}

Generative AI tools supported ideation, protocol and statistical design, adversarial review, experiment design, implementation, testing, evidence inspection, result interpretation, literature work, and manuscript editing. GPT and Codex assisted these activities. Claude interfaces supported review and provided the registered CLI execution layer. \texttt{deepseek-flash}, \texttt{deepseek-v4-flash}, and \texttt{deepseek-v4-pro} served as the audited agents and matched challengers in the respective studies. Image generation informed early teaser drafts; the publication teaser is native SVG artwork with vector export. Quantitative plots were generated with Matplotlib from frozen evidence. Executable evaluators and statistical code determine scores and decisions. The authors review the scientific claims, check numerical evidence and primary sources, test the code, and take responsibility for the final content and artifacts.

\bibliography{references}
\bibliographystyle{plainnat}

\clearpage
\appendix

\begin{figure}[t]
\centering
\includegraphics[width=\linewidth]{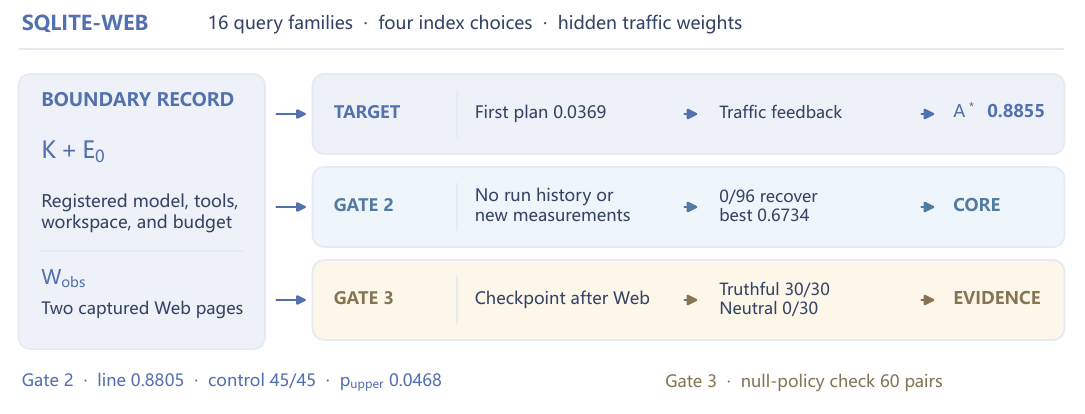}
\caption{One recorded SQLite-Web audit. The target captured two Web pages before its first experiment and used traffic feedback to select $A^*$. Gate 2 received those pages while $L^*$ and new task measurements were withheld. Gate 3 compared fresh feedback policies from a shared checkpoint after Web capture. The bottom strip reports controls and null-policy checks.}
\label{fig:sqlite-trace}
\end{figure}

\begin{figure}[t]
\centering
\includegraphics[width=\linewidth]{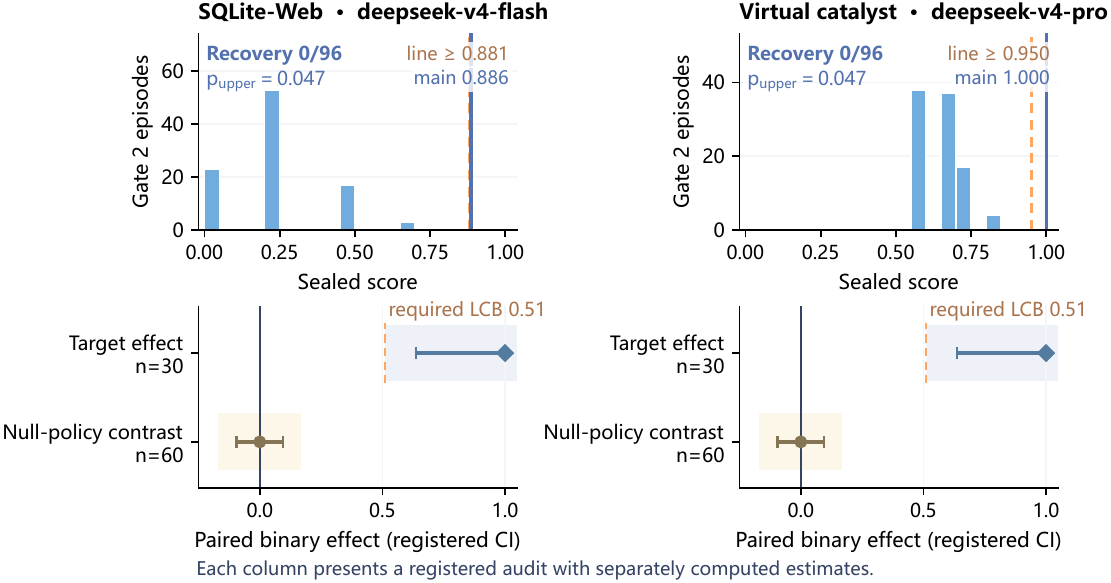}
\caption{Controlled three-gate audits in two domains and model scopes. The upper row shows all 96 Gate 2 scores per audit. The lower row shows registered 99\% intervals for the checkpoint-conditional feedback effect and independent null-policy checks.}
\label{fig:evidence}
\end{figure}

\begin{figure}[t]
\centering
\includegraphics[width=\linewidth]{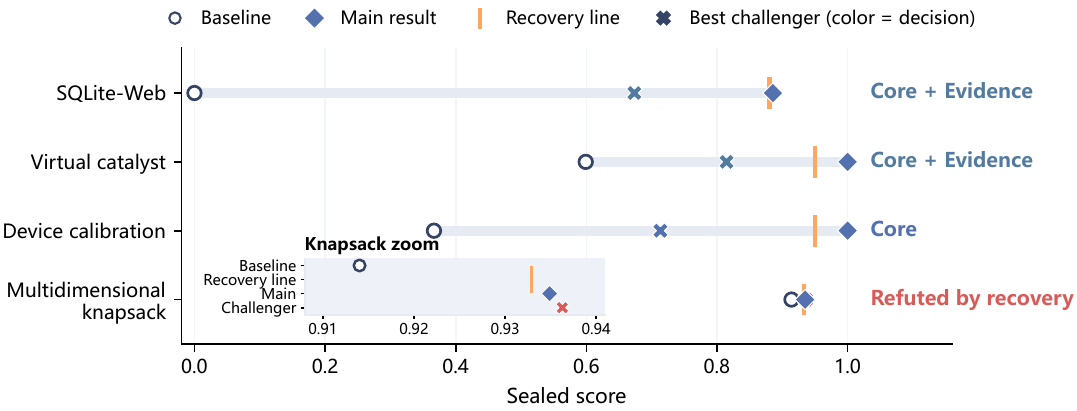}
\caption{The sealed-score geometry across four artifacts. The best no-lineage challenger stays below the registered recovery line in the three Core cases and crosses it in knapsack. The inset resolves the knapsack scores. The challenger may use any valid method.}
\label{fig:outcomes}
\end{figure}

\begin{figure}[t]
\centering
\includegraphics[width=0.96\linewidth]{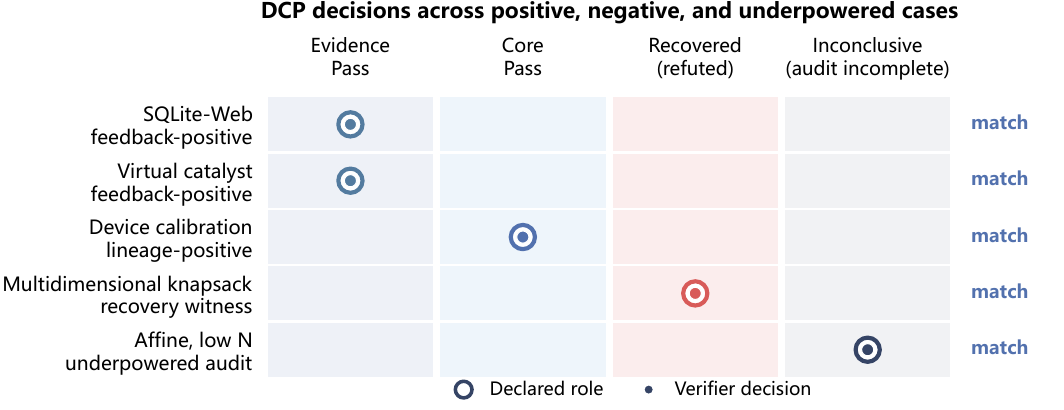}
\caption{The controlled calibration suite spans two Evidence positives, one constructed information-deficit Core positive, one developmental recovery refutation, and one intentionally undersampled audit. Rings show the declared calibration role and filled dots show the deterministic verifier decision.}
\label{fig:calibration}
\end{figure}

\section{Statistical Decision Details}

\subsection{Main result}

For paired private evaluation units, DCP computes differences between \(\astar\) and the baseline. The sampled-population profile uses a bounded empirical-Bernstein interval on the registered \([0,1]\) score scale. A claim over a fully enumerated finite set uses its exact mean. Each uncertainty calculation uses its registered sampling unit.

The value \(x\) is the registered evaluation mean of \(\astar\), selected before sealed scoring. The baseline must be valid and lie below the recovery region under its adjusted uncertainty check.

\subsection{Candidate decisions}

Each challenger candidate is scored on the same private units as \(\astar\). Let \(d_i\) be its paired score difference from \(\astar\). Candidate \(i\) is a confirmed hit when its adjusted lower confidence bound is at least \(-\epsilonv\). It is a confirmed miss when its adjusted upper confidence bound is below \(-\epsilonv\). Every other candidate is unresolved.

The family of candidate intervals shares a registered error budget. A simple profile assigns Bonferroni corrected error to every possible candidate slot, controlling false recovery across candidates that can change the decision. Equation~\ref{eq:zero} receives a separate error budget. Its prerequisites are zero qualified hits, resolved candidates, and a complete adequate control ledger.

The SQLite-Web and virtual catalyst audits allocated \(0.005\) each to main validation, candidate classification, positive-control adequacy, and neutral recovery, and \(0.01\) each to challenger recovery, the target feedback effect, and null-policy checks. The allocations total \(0.05\). Both used \(\rho=0.05\), 45 positive controls, and minimum control recall \(0.8\). Their feedback threshold was \(\delta_{\mathrm{evidence}}=0.34\) and decision buffer was \(\delta_{\mathrm{sham}}=0.17\). The real-data audits used the same per-gate error allocations with \(\delta_{\mathrm{evidence}}=0.20\) and \(\delta_{\mathrm{sham}}=0.10\).

When \(\alpha_{\mathrm{recovery}}=0.05\), 59 independent misses put the one-sided upper bound near \(0.05\). With \(\alpha_{\mathrm{recovery}}=0.025\), 72 misses are required. Device calibration allocated \(0.02\), required at least 77 misses, and registered 80 episodes. Each bound concerns one fresh episode from its registered model, policy, information, and budget distribution.

\subsection{Recovery witnesses and repeated search}

For independent Bernoulli episode outcomes with recovery probability \(p_B\), the probability of zero recoveries in \(n\) episodes is \((1-p_B)^n\). Thus a fixed-sample zero-hit audit reaches the registered bound when \(n\geq\lceil\log(\alpha_{\mathrm{recovery}})/\log(1-\rho)\rceil\). For the two controlled three-gate audits and all three real-data audits, \(\alpha_{\mathrm{recovery}}=0.01\) and \(\rho=0.05\) require 90 episodes; the registered count was 96. Candidate score classification and the other statistical gates retain their separate error allocations.

A witness establishes an available route, while the binomial bound quantifies the probability that a fresh episode produces any admissible recovery. DCP assigns them different decision roles. A campaign of \(k\) independent episodes from the same \(Q_B\) has recovery probability \(1-(1-p_B)^k\). A best-of-\(k\) production procedure therefore includes its full search and selection budget in one registered episode. Fresh Gate 3 pairs estimate the checkpoint-conditional feedback effect separately from selection of a successful target.

Additional episodes tighten a zero-hit bound and increase the chance of finding an existing rare route. For \(p_B=0.01\), 96 independent episodes find at least one witness with probability \(0.6190\); their probability of satisfying the zero-witness condition is \(0.3810\). Core uses this fixed-sample eligibility rule together with the bound threshold. A witness remains visible beside the bound and registered audit size. Table~\ref{tab:recovery-sensitivity} gives the exact-rule sensitivity. At \(\alpha_{\mathrm{recovery}}=0.01\), zero-hit audits need at least 44, 90, 228, and 459 episodes for \(\rho=0.10,0.05,0.02,0.01\), respectively.

\begin{table}[!ht]
\caption{Audit size and the two Gate 2 quantities. With \(\alpha_{\mathrm{recovery}}=0.01\), the upper bound conditions on zero hits. Witness detection assumes an illustrative fresh-episode recovery probability \(p_B=0.01\).}
\label{tab:recovery-sensitivity}
\centering
\small\rmfamily
\begin{tabular*}{\linewidth}{@{\extracolsep{\fill}}lrrrr@{}}
\toprule
\textbf{Independent episodes} & \textbf{30} & \textbf{96} & \textbf{192} & \textbf{384} \\
\midrule
Zero-hit recovery upper bound & 0.1423 & 0.0468 & 0.0237 & 0.0119 \\
Witness detection at \(p_B=0.01\) & 0.2603 & 0.6190 & 0.8548 & 0.9789 \\
\bottomrule
\end{tabular*}
\end{table}

\subsection{Feedback decisions}

The paired feedback estimator uses normalized utility differences or paired differences in recovery success, conditional on the frozen checkpoint. The utility and both policies are registered before the pairs start. The binary interval combines four exact one-sided Clopper--Pearson limits \citep{clopper1934use}. Let \(\delta_{\mathrm{evidence}}>0\) be the minimum effect and \(\delta_{\mathrm{sham}}>0\) the registered decision buffer. Support requires passing null-policy checks and \[ \operatorname{LCB}(\Delta_{\mathrm{feedback}}) \geq \delta_{\mathrm{evidence}}+\delta_{\mathrm{sham}}. \] The effect-below-threshold decision requires \[ \operatorname{UCB}(\Delta_{\mathrm{feedback}}) +\delta_{\mathrm{sham}} < \delta_{\mathrm{evidence}}. \] Remaining interval positions are inconclusive. The null interval describes equivalence on the registered known-answer family. The protocol quantifies neutral recovery with separate candidate checks and a finite-sample probability bound. Table~\ref{tab:pair-budget} gives the minimum supportive counts for the controlled audits' registered binary effect condition.

\begin{table}[htbp]
\caption{Registered binary feedback-effect decision boundary. Entries give the minimum truthful-only wins with zero neutral-only wins and all other pairs tied, using \(\alpha_{\mathrm{evidence}}=0.01\) and effect threshold \(\delta_{\mathrm{evidence}}+\delta_{\mathrm{sham}}=0.51\). ``No pass'' means that even all truthful-only wins fall short. Core, null-policy checks, and audit integrity have separate checks.}
\label{tab:pair-budget}
\centering
\small\rmfamily
\begin{tabular*}{\linewidth}{@{\extracolsep{\fill}}lccccc@{}}
\toprule
\textbf{Paired episodes} & \textbf{20} & \textbf{30} & \textbf{40} & \textbf{60} & \textbf{100} \\
\midrule
Minimum truthful-only wins & No pass & 28 & 35 & 48 & 71 \\
\bottomrule
\end{tabular*}
\end{table}

\section{Audit Validity}

Valid numerical interpretation requires both branches to implement the registered comparison. DCP records the following evidence.
\begin{itemize}
\item The registration is sealed before any private result or confirmatory null case is opened.
\item Model identity, interface, tools, information packet, and resource budget match their registered values.
\item Every started attempt remains in the ledger, including timeouts, invalid outputs, and infrastructure failures.
\item Candidate outputs are committed before private scores are returned.
\item The main output, baseline, and controls use one symmetric evaluator.
\item Each independent episode identifier corresponds to a committed draw record; its internal candidate opportunities share one statistical unit.
\item All Web responses given to the model are recorded byte for byte when Web access is enabled.
\end{itemize}

\paragraph{Classification versions.}
\label{app:failure-policy}
We use reporting identifiers for the real-data policies. \(F_0\) denotes each run's own frozen source policy, including its completion and reserved-channel checks. \(F_1\) names the recorded Ionosphere timeout amendment, and \(F_2\) names the Dry Bean regular-file amendment. Yacht retains \(F_0\); Ionosphere reports \(F_1\) alongside its \(F_0\) record; exploratory Dry Bean reports \(F_2\) alongside \(F_0\). These labels index archived policies and receipts. Table~\ref{tab:decision-provenance} gives amendment timing. Controlled tasks retain their task-specific registered rules.

\paragraph{Outcome failures and comparison integrity.} Under \(F_1\), a timeout with verified accepted-model activity and intact protected channels receives the registered failed-attempt utility. Under \(F_2\), an additional regular file containing public material leaves an existing valid candidate eligible for scoring; absent or invalid candidates receive failure utility. Wrong or unverified model identity, an executed forbidden tool, unauthorized information, and modified protected feedback remain incomplete-comparison conditions. Each source \(F_0\) retains its original checks and decision. Pre-action infrastructure faults follow the registered attrition rule, and every start remains in the ledger. Independent authority review and signatures govern formal issuance.

\begin{table}[!ht]
\caption{Decision vocabulary used throughout the paper. A pass denotes a local kernel decision. The reason field qualifies an Inconclusive decision.}
\label{tab:decision-labels}
\centering\small\rmfamily
\begin{tabularx}{\linewidth}{@{}l>{\raggedright\arraybackslash}X@{}}
\toprule
\textbf{Label} & \textbf{Meaning and recorded value} \\
\midrule
Pass & Registered numerical and audit checks pass; kernel value \texttt{certified}. \\
Refuted & A qualified recovery or confirmed failed claim condition; \texttt{refuted}. For Evidence, a confirmed effect-below-threshold finding maps here after Core, integrity, and null-policy prerequisites pass. \\
Inconclusive & Decision remains unresolved; \texttt{inconclusive}, with reason \texttt{audit\_incomplete} or \texttt{statistical\_uncertainty}. \\
Not evaluated & Optional grade omitted or blocked by a prerequisite; \texttt{not\_tested}. \\
\bottomrule
\end{tabularx}
\end{table}

\subsection{Episode budgets and sampling distributions}
\label{app:episode-budgets}

Table~\ref{tab:episode-budgets} specifies the recorded 96-episode audits, including exploratory Dry Bean. Budget \(B\) covers one challenger episode with one final-candidate opportunity; \(n\) counts audit repetitions. Each CLI session can contain several provider requests and file-tool calls, governed by its cost cap and timeout. Decoding uses provider defaults with effort \texttt{low}; commands impose no separate request-count or token cap. Target allowances are three \$0.50/360\,s sessions for SQLite-Web, two \$0.50/360\,s sessions for catalyst, and the same three-session caps as challengers for real-data tasks. Table entries distinguish these allowances from actual target use.

\begin{table}[!ht]
\caption{Challenger allowances, target consumption, and sampling scope. Every task registered \(n=96\) episodes with one final-candidate opportunity each. Target use lists sessions, reported USD, and summed CLI wall seconds, excluding evaluator and gateway time.}
\label{tab:episode-budgets}
\centering
\small\rmfamily
\renewcommand{\arraystretch}{1.12}
\setlength{\tabcolsep}{4pt}
\begin{tabularx}{\linewidth}{@{}>{\raggedright\arraybackslash}p{0.17\linewidth}>{\raggedright\arraybackslash}p{0.27\linewidth}>{\raggedright\arraybackslash}p{0.18\linewidth}>{\raggedright\arraybackslash}X@{}}
\toprule
\textbf{Task} & \textbf{Challenger budget \(B\)} & \textbf{Target use} & \textbf{Variation in \(Q_B\)} \\
\midrule
SQLite-Web &
1 session, \$1.50, 1{,}080\,s; 2 CPUs, 3\,GiB &
3 / 0.297 / 85.1 & Candidate and family aliases, row order, four prompt variants \\
\addlinespace[3pt]
Virtual catalyst &
2 sessions, each \$0.50 and 360\,s; 2 CPUs, 3\,GiB &
2 / 0.229 / 93.0 & Public nonce and five-control starting recipe \\
\addlinespace[3pt]
Yacht &
Up to 3 sessions, each \$1.00 and 240\,s; 2 CPUs, 2\,GiB &
2 / 0.352 / 91.6 & Fixed task; fresh provider sampling \\
\addlinespace[3pt]
Ionosphere &
Up to 3 sessions, each \$1.00 and 240\,s; 2 CPUs, 2\,GiB &
2 / 0.280 / 78.4 & Fixed task; fresh provider sampling \\
\addlinespace[3pt]
Dry Bean Web (R) &
Up to 3 sessions, each \$1.00 and 360\,s; 2 CPUs, 2\,GiB &
3 / 0.483 / 121.8 & Fixed task and Web packet; fresh provider sampling \\
\bottomrule
\end{tabularx}
\end{table}

\paragraph{Fixed opportunity and stopping rules.} All episodes use the registered model, public task, initial evidence, evaluator, thresholds, and stateless CLI interface. SQLite challengers receive all target-observed Web bytes and commit one index plan in one session. Catalyst challengers choose an assay anchor in the first session and commit a recipe in the second, with a registered no-signal assay between sessions. The nonce maps each starting control independently to \(\{0,1,2,4,5,6,7\}\). Real-data challengers may select eight masks and receive an empty measurement map; the evaluator uses their final submitted table with the two public anchors. Their action loop ends on \texttt{finish}, a session failure, or the three-session cap. Cost exhaustion and timeout terminate each session. Intermediate edits remain within the same episode, and final scoring opens after all candidate commitments.

\paragraph{Provider model identifiers.} Assistant payloads report \texttt{deepseek-v4-flash} for SQLite-Web, device calibration, and knapsack, \texttt{deepseek-v4-pro} for catalyst, and \texttt{deepseek-flash} for Yacht, Ionosphere, and Dry Bean. The configured DeepSeek endpoint routes the CLI model slot \texttt{claude-sonnet-4-6} to the first and third backends; this slot is a client-side alias. Catalyst requests \texttt{deepseek-v4-pro} directly. Identity checks use the provider's assistant field. The identifiers, frozen endpoints, and collection windows define the model scope; the provider supplies no weight-checkpoint revision for \texttt{deepseek-flash}. Short figure labels use v4-flash, v4-pro, and flash.

\paragraph{Collection windows.} All real-data model sessions were collected on 25 September 2026. Recorded session timestamps, rounded to seconds in UTC, span 10:58:42 to 15:26:16 for Yacht, 11:13:45 to 14:01:20 for Ionosphere, and 16:49:28 to 21:10:18 for Dry Bean. These windows run from the first target-session start through the last audit-session completion, including all started branches.

\paragraph{Positive-control inputs.} SQLite-Web controls receive the known optimal plan, an alias-mapping test, and a captured-Web content marker. Catalyst controls receive a complete informative assay for a fresh reaction and must submit its optimum. Device controls receive every hidden gain in episode-local coordinates. Yacht, Ionosphere, and Dry Bean controls receive a task-neutral marker in \texttt{CANARY.txt}, copy it exactly, and submit a finish action. These checks respectively assess supplied-solution submission, assay-to-recipe execution, calibration submission, and file-channel operation. Challenger search performance is evaluated through the recorded recovery attempts and their stated \(Q_B\).

\paragraph{Information and sampling scope.} Controlled sessions expose Read and Edit tools; real-data sessions expose Read, Write, Edit, Glob, and Grep. Shell and native Web tools are disabled. Yacht, Ionosphere, and catalyst use no Web. Dry Bean permits one host-mediated UCI request with a 20\,s timeout, a 2{,}000{,}000-byte raw limit, and a 300{,}000-byte delivery limit; Gate 2 uses the recorded response. All 96 initial workspace manifests and first-prompt receipts match within each real-data task. Their independent, identically distributed episode model concerns repeated provider sampling conditional on that fixed configuration. Draw receipts identify started episodes; the recorded real-data commands leave the provider's random seed unspecified. The resulting \(p_B\) bound applies to each task's stated generation process and budget.

\section{Audit Case Details}
\label{sec:case-details}

SQLite-Web denotes the event-traffic profile with two captured documentation pages; SQLite with Web disabled is a separate earlier Core audit, and the e-commerce profile supplies the separate parent and Gate 3 addendum in Appendix~\ref{app:audit-record}.

\begin{table}[!ht]
\caption{Controlled audit decisions. Gate 1 shows baseline-to-target utility; T/N gives truthful/neutral recoveries. Zero-hit bounds are \(0.0468\) for 96 episodes and \(0.0477\) for 80, with respective recovery error budgets \(0.01\) and \(0.02\). Knapsack supplies two developmental witnesses.}
\label{tab:controlled-values}
\centering
\small\rmfamily
\setlength{\tabcolsep}{3pt}
\renewcommand{\arraystretch}{1.1}
\begin{tabular*}{\linewidth}{@{\extracolsep{\fill}}lrrrrccl@{}}
\toprule
\textbf{Case} & \shortstack{\textbf{Gate 1}\\\textbf{score}} &
\shortstack{\textbf{Best}\\\textbf{challenger}} &
\shortstack{\textbf{Recovery}\\\textbf{line}} &
\textbf{Recovery} & \textbf{Controls} &
\shortstack{\textbf{Gate 3}\\\textbf{T/N}} & \textbf{Decision} \\
\midrule
\shortstack[l]{SQLite-Web\\v4-flash} & \shortstack[r]{0.0000\\\(\to 0.8855\)} &
0.6734 & 0.8805 & 0/96 & 45/45 &
\shortstack{30/30, 0/30\\null 60, pass} & \shortstack[l]{\textbf{Core +}\\\textbf{Evidence}} \\
\addlinespace[2pt]
\shortstack[l]{Virtual catalyst\\v4-pro} & \shortstack[r]{0.5990\\\(\to 1.0000\)} &
0.8146 & 0.9500 & 0/96 & 45/45 &
\shortstack{30/30, 0/30\\null 60, pass} & \shortstack[l]{\textbf{Core +}\\\textbf{Evidence}} \\
\addlinespace[2pt]
\shortstack[l]{Device calibration\\v4-flash} & \shortstack[r]{0.3667\\\(\to 1.0000\)} &
0.7133 & 0.9500 & 0/80 & 60/60 & Not run & \textbf{Core} \\
\addlinespace[2pt]
\shortstack[l]{Knapsack\\v4-flash, dev.} & \shortstack[r]{0.9140\\\(\to 0.9349\)} &
0.9363 & 0.9329 & Found & n/a & Not run & \textbf{Refuted} \\
\bottomrule
\end{tabular*}
\end{table}

\subsection{SQLite with captured Web access}

The registered assistant payload was \texttt{deepseek-v4-flash}. The main score was \(0.8855470\), the baseline score was \(0\), and the recovery line was \(0.8805470\). All 96 matched challenger episodes were valid confirmed misses. Their scores took four values between \(0.0368978\) and \(0.6733847\). The zero-hit upper bound was \(1-0.01^{1/96}=0.0468381\). All 45 positive controls succeeded, giving a one-sided recall lower bound of \(0.8889268\). Gate 1 used the exact enumerated finite workload and required a gain of \(\delta_{\min}=0.5\).

The gateway recorded two HTTPS responses from \texttt{sqlite.org}. Gate 2 received the normalized model-visible text and content hash for every response. The target feedback test used 30 fresh pairs. Truthful and neutral recovery counts were \(30/30\) and \(0/30\). The paired exact 99\% interval was \([0.6379275,1]\). In 60 fresh null pairs, both arms succeeded in every pair, giving a difference of \(0\) and an exact 99\% interval of \([-0.0950339,0.0950339]\).

The machine-readable record is \path{artifacts/dcp_sqlite_event_web_fullflow_confirmatory_v1_certificate/certificate.json}. Its identifier begins with \texttt{b3f0af47968ba2cb}. The self-contained publication Bundle containing the same certificate is \path{examples/audits/sqlite-web} and its identifier begins with \texttt{eff6cf27afa58da5}. All SQLite-Web decisions and Gate 3 numbers in this paper are read from this certificate. Appendix~\ref{app:audit-record} records its verifier-versioned provenance.

Confirmatory collection used 504 model sessions and \(61.13\) USD. Including three interface preflight sessions gives 507 sessions and \(61.17\) USD.

\subsection{Virtual catalyst optimization}

The registered model identity was \texttt{deepseek-v4-pro}. The main score was \(1.0\), the baseline score was \(0.5989989\), and the recovery line was \(0.95\). Gate 1 required \(\delta_{\min}=0.35\), with recovery tolerance \(\epsilonv=0.05\). All 96 matched challenger episodes were valid confirmed misses, and the best score was \(0.814649\). The zero-hit upper bound was \(1-0.01^{1/96}=0.0468381\). All 45 positive controls succeeded, giving a one-sided recall lower bound of \(0.8889268\).

The target feedback study used 30 fresh pairs. Truthful and neutral recovery counts were \(30/30\) and \(0/30\). The paired exact 99\% interval was \([0.6379275,1]\). In 60 fresh null pairs, both arms had equal outcomes in every pair, giving a difference of \(0\) and an exact 99\% interval of \([-0.0950339,0.0950339]\).

Replay uses \path{examples/audits/virtual-catalyst}. Its identifier begins with \texttt{fa17603f2184bb7b}.

Confirmatory collection used 419 model sessions and \(54.27\) USD. Development and confirmatory collection together used 435 sessions and \(56.40\) USD. Both source and replay records accompany the supplement.

\subsection{SQLite with Web disabled}

This secondary Core audit predates the Web-enabled case and is retained for direct replay at \path{examples/audits/sqlite-core}. The main score was \(0.9432886\), the baseline was \(0\), and the recovery line was \(0.9382886\). All 80 challenger episodes were valid confirmed misses. The best challenger scored \(0.7947267\). The zero-hit upper bound was \(1-0.02^{1/80}=0.0477239\). All 45 positive controls succeeded, giving a one-sided recall lower bound of \(0.8889268\). Gate 1 used the exact enumerated finite workload and required \(\delta_{\min}=0.5\). The evidence bundle identifier is \texttt{3690bebc8b3ac882}. The 130 recorded model sessions cost \(25.56\) USD.

\subsection{Device calibration}

The public replay Bundle is \path{examples/audits/device-calibration-core}. The main score was \(1.0\), the baseline score was \(110/300=0.3667\), and the recovery line was \(0.95\). All 80 challenger episodes were valid confirmed misses. The best score was \(214/300=0.7133\). The zero hit upper bound was \(1-0.02^{1/80}=0.0477\). All 60 positive controls succeeded, producing a one sided Clopper Pearson lower bound of \(0.9155\). Gate 1 used the exact enumerated 300-item set and required \(\delta_{\min}=0.5\). Its identifier begins with \texttt{dff7b57b062362e0}. The recorded provider cost was \(19.08\) USD.

\subsection{Multidimensional knapsack}

The main score was \(0.9348968\), the baseline score was \(0.9140206\), and the recovery line was \(0.9328968\). Two candidate opportunities inside one challenger episode produced scores \(0.9362833\) and \(0.9355607\). These qualified recovery witnesses trigger the Core veto. The anonymous supplement provides a six-file challenger input manifest whose reconstructed digest matches both prestart records in the frozen event chain. The packet marks task feedback unavailable and contains no earlier candidate score. The record supports a constructive recovery decision. On the deterministic finite sealed set, the main gain was \(0.0208762\), above the registered \(\delta_{\min}=0.01\).

\subsection{Development checks retained as inconclusive}

An affine parity run exercised all three gates with two challenger episodes, two target pairs, and two null pairs. The observed main score and feedback contrast were both \(1.0\), while the positive-control recall lower bound was \(0.0707\). The verifier returned an incomplete audit at control adequacy. An active automaton run showed a \(1.0\) main score against \(0.5\) no-feedback outputs and remained incomplete in challenger adequacy and paired interface records. Earlier single-response induction tests measure \(E_0\)-dependence because their labeled examples were available at the start. These developmental records are reported separately from the primary audits.

\section{Real-Data Workflow Records}
\label{app:real-workflows}

The audited artifact in these workflows is a completed response surface for 16 executable configurations. For a submitted prediction table \(\hat u\) and an authorized observation map \(O\), the evaluator applies the fixed rule
\begin{equation}
    \astar_m=C(\hat u,O)_m=
    \begin{cases}
        O(m), & m\in\operatorname{dom}(O),\\
        \hat u_m, & m\notin\operatorname{dom}(O).
    \end{cases}
    \label{eq:surface-completion}
\end{equation}
Its registered outcome measures recovery of 32 adjacent-configuration utility effects. Clustering utility is NMI against private labels, Yacht uses \(1/(1+\mathrm{MAE}/\mathrm{IQR}_{\mathrm{train}})\), and Ionosphere uses balanced accuracy. The two public anchors are \(E_0\). Eight utilities returned by the truthful measurement action enter \(\lineage\) and expand \(O\) to ten entries; the other six remain private through artifact commitment. Challenger and neutral observations contain only the two anchors. Final effect-recovery scores are revealed after candidate generation closes. Unmeasured-six MAE summarizes prediction error on configurations whose utilities stayed private.

\paragraph{Baseline and insertion-only comparison.} The registered baseline preserves the two public anchors and assigns their mean to the other 14 entries. The offline insertion-only diagnostic additionally inserts the target's eight frozen measurements and leaves its six remaining entries at that same mean. Table~\ref{tab:insertion-only} reports this fixed-batch comparison. Insertion alone clears the registered \(\delta_{\min}=0.20\) gain for Yacht and Ionosphere. Agent predictions further improve their effect-reconstruction scores by \(0.4691\) and \(0.2289\). Gate 1 measures the full acquisition-and-completion gain. The tolerance remains \(\epsilonv=0.05\), with \(\alpha_{\mathrm{recovery}}=0.01\) and \(\rho=0.05\). Gate 3 requires a feedback lower bound of \(0.30\) and a null interval inside \([-0.10,0.10]\).

\begin{table}[!ht]
\caption{Offline completion diagnostic on the frozen target measurement sets. Insertion-only adds eight observed utilities to the two-anchor baseline and uses the anchor mean for the remaining six entries. The final column measures the additional score from the submitted predictions at fixed observations. The registered Gate 1 baseline remains the two-anchor column.}
\label{tab:insertion-only}
\centering\small\rmfamily
\begin{tabular*}{\linewidth}{@{\extracolsep{\fill}}lrrrr@{}}
\toprule
\textbf{Task} & \textbf{Two anchors} & \textbf{Insertion-only} & \textbf{Completed target} & \textbf{Added score} \\
\midrule
Yacht & 0.0000 & 0.2986 & 0.7677 & +0.4691 \\
Ionosphere & 0.0988 & 0.4263 & 0.6551 & +0.2289 \\
Dry Bean (R) & 0.0000 & 0.1439 & 0.7124 & +0.5685 \\
\bottomrule
\end{tabular*}
\end{table}

\paragraph{Registered intervention.} The adapter selects checkpoint 0, containing the public task, two anchors, and initial prediction table, before any eight-mask selection. In Dry Bean this state also precedes Web capture; both fresh arms first replay the same frozen UCI response. Each arm then chooses its own eight-mask batch. Truthful replies use schema \texttt{wwb\_feedback\_v1}, status \texttt{measured}, and an eight-entry measurement map. Neutral replies use the same schema, status \texttt{no-measurement}, and an empty map. The agent subsequently submits its prediction table, and Equation~\ref{eq:surface-completion} inserts the authorized values. The Gate 3 estimand is the end-to-end recovery effect on this completed data product. It includes direct insertion of eight measured values and any improvement in predictions for the remaining six configurations. Even with unchanged residual predictions, inserting measured values can improve the score. The known-answer null family compares empty-map policies labelled \texttt{no-measurement} and \texttt{reference-null}.

\subsection{Exploratory Dry Bean Web response surface}
\label{app:dry-bean}

Dry Bean uses bean-image features and seven private labels for clustering \citep{uci2020drybean}. Its four switches control signed-log features, scaling, 95\% principal-component projection, and one versus ten KMeans initializations. Utility is normalized mutual information (NMI). This exploratory record complements the primary Yacht and Ionosphere audits.

\begin{figure}[!ht]
\centering
\includegraphics[width=\linewidth]{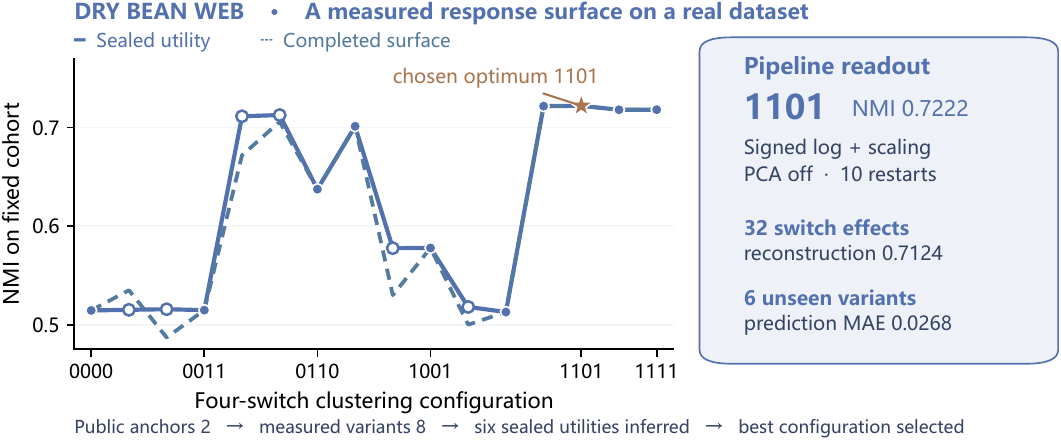}
\caption{Exploratory Dry Bean response surface. Ten observed utilities and six predictions form the scored data product. Pipeline selection is an accompanying readout. Table~\ref{tab:decision-provenance} records the original and revised-policy kernel decisions.}
\label{fig:dry-bean-exploratory}
\end{figure}

The captured UCI page has snapshot identifier \texttt{54e2ea93} and is replayed byte for byte. Figure~\ref{fig:dry-bean-exploratory} compares the completed surface with all 16 reference utilities. Full-grid MAE is \(0.0100\), and unmeasured-six MAE is \(0.0268\). Reconstructed mean effects for signed log, scaling, PCA, and ten restarts are \((+0.0291,+0.1782,-0.0237,+0.0259)\); sealed values are \((+0.0304,+0.1747,-0.0271,+0.0075)\). Mean absolute true edge effect \(0.0605176846\) and edge error \(0.0174069010\) give reconstruction score \(0.7124\). Measured configuration \texttt{1101} attains the highest pipeline NMI, \(0.7222\), as an accompanying selection result. The final analysis bundle at \path{examples/audits/dry-bean-web}, identifier \texttt{d5c04462cdd1ffb9}, returns Core Pass and Evidence Pass under the revised analysis policy. Reported model cost is \(48.39\) USD.

\paragraph{Exploratory analysis scope.} The source record has an incomplete challenger-contract attestation because episode 78 created an extra regular file in the reserved channel directory. The file contained a public request identifier; recorded protected observations and instructions stayed unchanged. The revised policy treats this file addition as an ordinary workspace error and scores the existing candidate, preserving all 96 slots. That policy was recorded after Gate 2 scoring, so its Core and Evidence passes describe an exploratory reanalysis. The original Core Inconclusive and dependent Evidence Not evaluated decisions remain in Table~\ref{tab:decision-provenance}. This case supplies response-surface and feedback observations within that stated scope.

\paragraph{Paired feedback statistics.} The separately collected Gate 3 record contains 30 target pairs and 60 null pairs. Its 27 truthful-only recoveries give effect \(0.90\) and 99\% interval \([0.4772,0.9910]\); all null contrasts are zero, with interval \([-0.0950,0.0950]\). These observations are present in both bundles. They quantify the end-to-end feedback effect independently of the original certificate's short-circuited Evidence field. The revised-policy Evidence decision inherits the reanalysis scope of Core.

\subsection{Yacht and Ionosphere}

Yacht's target score was \(0.7677\), against a baseline of \(0\), and its best challenger scored \(0.4641\) below the \(0.7177\) recovery line. Its full-audit bundle at \path{examples/audits/yacht} has identifier prefix \texttt{72e6c50143839ba8}. The target feedback effect interval included zero. One null branch modified the protected observation during calibration pair 25, leaving \(59\) analyzable null pairs and Evidence Inconclusive with an incomplete-audit label. The recorded model cost was \(39.23\) USD.

Ionosphere's target score was \(0.6551\), and its best challenger scored \(0.4171\) below the \(0.6051\) recovery line. Its versioned full-audit bundle at \path{examples/audits/ionosphere} has identifier prefix \texttt{017a4fd527476dbd}. All 60 null pairs were analyzed, with one nonzero contrast. Their equivalence interval \([-0.0950,0.1289]\) crossed the registered \(\pm0.10\) band; the target effect lower bound \(0.0944\) stayed below \(0.30\). An accepted-model timeout was classified as a failed branch in the versioned record, and the Evidence verdict remained inconclusive. The recorded model cost was \(32.98\) USD. Both retain local Core Pass with zero recoveries in 96 attempts and \(45/45\) file-channel controls.

\section{Audit Record}
\label{app:audit-record}

\begin{table}[!ht]
\caption{Decision provenance and classification versions. P denotes Pass, I denotes Inconclusive, and N denotes Not evaluated. Decision pairs list Core/Evidence. Policy entries give source-to-report versions defined in Appendix~\ref{app:failure-policy}. Appendix~\ref{app:audit-record} specifies cross-bundle replay coverage.}
\label{tab:decision-provenance}
\centering\small\rmfamily
\renewcommand{\arraystretch}{1.15}
\setlength{\tabcolsep}{4pt}
\begin{tabularx}{\linewidth}{@{}>{\raggedright\arraybackslash}p{0.16\linewidth}cc>{\raggedright\arraybackslash}p{0.13\linewidth}>{\raggedright\arraybackslash}X@{}}
\toprule
\textbf{Record} & \textbf{Source} & \textbf{Reported} & \textbf{Policy} & \textbf{Version basis} \\
\midrule
SQLite-Web & P/I & P/P & Task policy retained & Corrected read-before-first-write implementation; all 120 saved null branches replayed. Source Evidence reason was audit incomplete. \\
\addlinespace[3pt]
Virtual catalyst & P/P & P/P & Task policy retained & Original registration and decision retained. \\
\addlinespace[3pt]
Yacht & P/I & P/I & \(F_0\to F_0\) & Original decision retained. A changed protected observation leaves the null-policy comparison audit incomplete. \\
\addlinespace[3pt]
Ionosphere & P/I & P/I & \(F_0\to F_1\) & Accepted-model timeout amendment recorded after start and before Gate 3 scoring. Both versions retain statistical uncertainty; Core is unchanged. \\
\addlinespace[3pt]
Dry Bean (R) & I/N & P/P (R) & \(F_0\to F_2\) & Regular-file amendment recorded after Gate 2 scoring. Revised passes are exploratory; source Core remains audit incomplete. \\
\bottomrule
\end{tabularx}
\end{table}

Each audit record binds the registered claim and information boundary to its final output, sealed scores, challenger contract, attempt ledger, controls, intervals, and decision. Feedback records add paired branches and null-policy checks. The verifier replays these records offline. Local results retain \path{formal_certificate_issued=false}; independent provenance verification and countersigning govern formal issuance. Historical priority remains \texttt{not\_assessed}.

\paragraph{Cross-bundle replay.} The current kernel replays all 11 supplied audit bundles and the linked e-commerce addendum, including the archived source versions, with every saved decision reproduced. Raw SQLite checks cover the event-traffic audit, the e-commerce parent, and its addendum under their recorded path rules. The read-ordering implementation correction changes only the event-traffic Evidence decision from Inconclusive to Pass. Ionosphere retains its decisions under \(F_1\), and Dry Bean's exploratory changes are separately attributed to policy \(F_2\).

\paragraph{Separate registered SQLite feedback record.} The earlier e-commerce parent \texttt{1f9d3142} yielded Core Pass and Evidence Inconclusive. Its registered Gate 3 addendum \texttt{210b680c} started 22 target pairs and 45 null pairs with \(\alpha_{\mathrm{evidence}}=\alpha_{\mathrm{sham}}=0.01\). Twenty-one truthful branches reached the target, one had a model failure, and all 22 neutral outputs stayed below the recovery line. Raw replay identifies an unfinished, unverified experiment session in reference-null pair 28 and relative-path reads outside the frozen exact-path convention in pair 32. The ordering correction leaves both conditions unresolved, preserving Evidence Inconclusive. The parent and addendum use the recorded exact-path convention; event-traffic uses its separately frozen workspace-normalized path policy.

\paragraph{Verifier-versioned event-traffic certificate.} An earlier ordering verifier marked two sequential-write sham branches incomplete in frozen event-traffic bundle \texttt{2889e78f}. Offline recertification applied the registered exact-read-before-first-write rule, replayed subsequent writes in order, and validated all 120 sham branches. Raw sessions, registration, Gate~1 and Gate~2 inputs, target pairs, thresholds, and error allocations stayed fixed. The canonical certificate \texttt{b3f0af47} is packaged in publication bundle \texttt{eff6cf27}; its recertification receipt is in the supplement.

\end{document}